\documentclass[twocolumn,amsmath,amssymb,longbibliography,superscriptaddress]{revtex4-1}

\usepackage{graphicx}
\usepackage{dcolumn}
\usepackage{bm}

\usepackage{amsfonts} 
\usepackage{mathtools}

\usepackage{esdiff}
\usepackage{braket}

\usepackage[english]{babel}

\usepackage[T1]{fontenc}
\usepackage[utf8]{inputenc} 
\usepackage{lmodern}
\usepackage{microtype}

\usepackage{color}
\usepackage[colorlinks=true,allcolors=blue]{hyperref}

\newcommand{\ii}{\mathrm{i}} 
\newcommand{\ee}{\mathrm{e}} 
\newcommand{\var}[1]{{(\Delta #1)^2}}
\newcommand{\rb}{$\prescript{87}{}{\text{Rb}}$ }

\begin{document}
\title{Active SU(1,1) atom interferometry}

\date{\today}

\author{D. Linnemann}
\email{activeinterferometer@matterwave.de}
\author{J. Schulz}
\author{W. Muessel}
\author{P. Kunkel}
\author{M. Prüfer}
\author{A. Frölian}
\author{H. Strobel}
\author{M. K. Oberthaler}
\affiliation{Kirchhoff-Institut f\"ur Physik, Universit\"at Heidelberg, Im Neuenheimer Feld 227, 69120 Heidelberg, Germany.}

\begin{abstract}
Active interferometers use amplifying elements for beam splitting and recombination. We experimentally implement such a device by using spin exchange in a Bose-Einstein condensate. The two interferometry modes are initially empty spin states that get spontaneously populated in the process of parametric amplification. This nonlinear mechanism scatters atoms into both modes in a pairwise fashion and generates a nonclassical state. Finally, a matched second period of spin exchange is performed that nonlinearly amplifies the output signal and maps the phase onto readily detectable first moments. Depending on the accumulated phase this nonlinear readout can reverse the initial dynamics and deamplify the entangled state back to empty spin states. This sequence is described in the framework of SU(1,1) mode transformations and compared to the SU(2) angular momentum description of passive interferometers. 

\vspace{10px} \textbf{Published in} \href{https://doi.org/10.1088/2058-9565/aa802c} {Quantum Science and Technology {\bf 2}, 044009 (2017)}

\end{abstract}

\maketitle
\begin{figure*}
\includegraphics{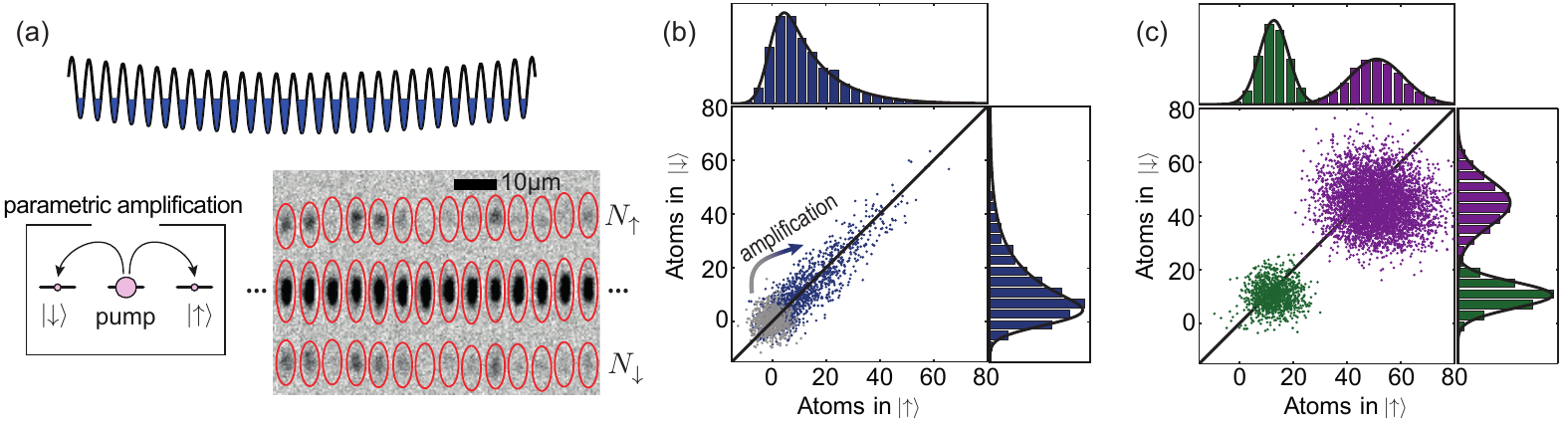}
\caption{\textbf{Experimental realisation of parametric amplification} \textbf{(a)} We use spin mixing among three hyperfine levels in a spinor Bose-Einstein condensate. The atoms are trapped in a one-dimensional standing wave potential with additional transverse harmonic confinement. Atom numbers of the three spin states are obtained by state and lattice site resolving absorption imaging. A typical absorption image with counting regions indicated by ellipses is shown. \textbf{(b)} Atom number correlations of the two modes $\ket{\uparrow}$ and $\ket{\downarrow}$ as a result of parametric amplification. The spread orthogonal to the diagonal is dominated by detection noise. The initial vacuum state is shown in grey. The mode populations after parametric amplification resemble a thermal-like state as shown by the respective histogram. \textbf{(c)} Atom number correlations after passive beam splitting. The green points depict a state with similar mean atom number to the state produced by parametric amplification. The mode populations are Gaussian as indicated by the histograms. Their width is a combination of detection noise and atomic shot noise. The contribution of the latter dominates for state with larger mean atom number (purple).}
\end{figure*}

\section{Introduction}

This year marks the $125$th anniversary of the Mach-Zehnder interferometer \cite{Mach1892}. When transferring the basic idea of two path interference from light to massive particles, Feynman famously concluded that this is a ``phenomenon which has in it the heart of quantum mechanics \cite{Feynman1963}.'' Nowadays, the atomic analogue of Mach-Zehnder interferometry is an indispensable instrument in fundamental research as well as technology \cite{Cronin2009, Degen2016}. Its prime example is the atomic clock \cite{Santarelli1999}. 

These routinely employed interferometers use passive beam splitters during which a conserved number of particles is redistributed among two modes. In contrast, an active interferometer uses amplifying elements for beam splitting and recombination. Consequently, the total particle number in the corresponding interferometry modes becomes subject to change. Such an interferometer is described within the framework of the SU(1,1) group \cite{Yurke1986}. Due to the intrinsic amplification they provide benefits for detecting phase shifts of weak signals. Furthermore, the exploitation of entanglement \cite{Giovannetti2006} generated during quantum amplification allows reaching phase sensitivities better than those of classical passive interferometers \cite{Linnemann2016}. 

Originally devised for optics \cite{Yurke1986}, the essential idea of using two nonlinear processes with phase accumulation in-between has been realised with microwave circuit QED \cite{Flurin2012}, spinor Bose-Einstein condensates \cite{Hoang2013}, nonlinear optical fibres \cite{Vered2015}, Raman scattering \cite{Chen2015} and four-wave mixing \cite{Pooser2009} in vapour gases. Improved phase sensitivities have been demonstrated in optics \cite{Hudelist2014} and with atoms \cite{Linnemann2016}. 

The centrepiece of SU(1,1) interferometry is the process of parametric amplification which is used as an active beam splitter. We devote the first part of this manuscript to detail its experimental implementation. We then present a theoretical framework of SU(1,1) transformations while contrasting it to the well known passive SU(2) interferometer. Finally, experimental results are discussed. The findings reported on in this manuscript build on the work of \cite{Linnemann2016}. 

\section{Experimental system: Spinor Bose-Einstein condensate}

Bose-Einstein condensates are a well established system to study nonlinear spin dynamics \cite{Stamper-Kurn2013}, interferometry \cite{Cronin2009}, sensing \cite{Degen2016}, quantum metrology \cite{Pezze2016}, and quantum information \cite{Braunstein2005}. 

We experimentally realise the SU(1,1) transformations within the spin degree of freedom of a \rb Bose-Einstein condensate. For this, the atoms are tightly trapped in an optical standing wave potential such that their external degrees of freedom are frozen out. In this single-spatial mode, dynamics is restricted to the spin degree of freedom. 

For parametric amplification we employ spin exchange \cite{Stamper-Kurn2013} among three hyperfine levels. For this purpose a large population of $N_0\approx500$ atoms in $m_F=0$ acts as a particle reservoir (pump mode). Mediated by coherent collisions of two pump atoms, the initially empty modes $m_F=1\equiv\ket{\uparrow}$ and $m_F=-1\equiv\ket{\downarrow}$ get populated in a pairwise manner (Fig 1a) such that the total $m_F$ remains conserved. This process is described by the Hamiltonian $\mathcal{H}=\hbar \kappa (\hat{a}_\downarrow^\dagger\hat{a}_\uparrow^\dagger+\text{h.c.})$ \cite{Linnemann2016}. Here $\hat{a}_\uparrow^\dagger$ ($\hat{a}_\downarrow^\dagger$) are the creation operators for mode $\ket{\uparrow}$ ($\ket{\downarrow}$) and $2\pi\hbar$ denotes Planck's constant. The effective nonlinear coupling strength $\kappa=gN_0$ is enhanced by the number of pump atoms and arises microscopically as a result of different s-wave collision channels \cite{Widera2006} which is parametrised by the scattering length difference $\Delta a$. Then $g \propto \Delta a \int \! \mathrm{d}^3x |\Psi(\bm{x})|^4$ where $\Psi(\bm{x})$ is the condensates spatial mode function. This Hamiltonian remains valid as long as the pump mode acts as an unlimited reservoir and therefore exceeds all other mode populations by far. In this undepleted pump approximation, both, the sign and the magnitude of the nonlinear coupling strength $\kappa$ can be controlled via the pump mode: While its population defines the nonlinearity's magnitude, the phase of the pump mode controls the phase of $\kappa$. In practice, we transfer the pump atoms into an ancilla spin state that does not participate in parametric amplification to change the nonlinearity's magnitude. We also use such state transfers to imprint a dynamical phase onto the pump mode. 

Atomic populations of the individual spin components are detected via absorption imaging (Fig 1a) with an uncertainty (one s.d.) of $4$ atoms \cite{Muessel2013}. All lattice sites are independent with respect to the spin dynamics and used to increase the statistical sample size. For quantitative analysis we postselect experimental runs with $450-500$ atoms per lattice site such that the nonlinearity $\kappa$ is fixed on the 10\% level. To accurately estimate mean values and their variances we repeat the experiment typically a few hundred times.

All experiments start with empty modes $\ket{\uparrow}$ and $\ket{\downarrow}$ (vacuum state). The characteristic mode correlations subsequently generated by parametric amplification become manifest in Fig 1b. Here, the single-shot spin populations line up on the diagonal $N_\uparrow=N_\downarrow$. Only small deviations from this line occur and are mainly caused by detection infidelities. This becomes apparent when comparing this width to the measured initial vacuum state (grey), whose isotropic extension is exclusively due to detection noise. Along the diagonal, however, the quantum state shows excess noise, i.e. fluctuations in total atom number $N_\uparrow+N_\downarrow$. The population of the modes shows distinctive skewed distributions as witnessed by the histograms. The solid line is a fit to the theoretical expectation of a thermal-like distribution which also takes into account detection noise by Gaussian convolution. 

A state with similar mean atom number produced by a passive beam splitter is shown in panel c (green). Here a resonant three-level rf-pulse couples the pump mode to $\ket{\uparrow}$ and $\ket{\downarrow}$, which thereby populates these modes. The resulting state is much more concentrated around its mean atom number and its fluctuations are isotropic. In this case, the state's extension is caused by both, detection noise \emph{and atomic shot noise} as a consequence of Poissonian statistics. The fluctuations orthogonal to the diagonal, i.e. in particle number difference $N_\uparrow-N_\downarrow$, are therefore enlarged compared to parametric amplification. This becomes even more pronounced when considering a state with larger mean atom number (purple). 

\section{Theoretical description}

\begin{figure*}
\includegraphics{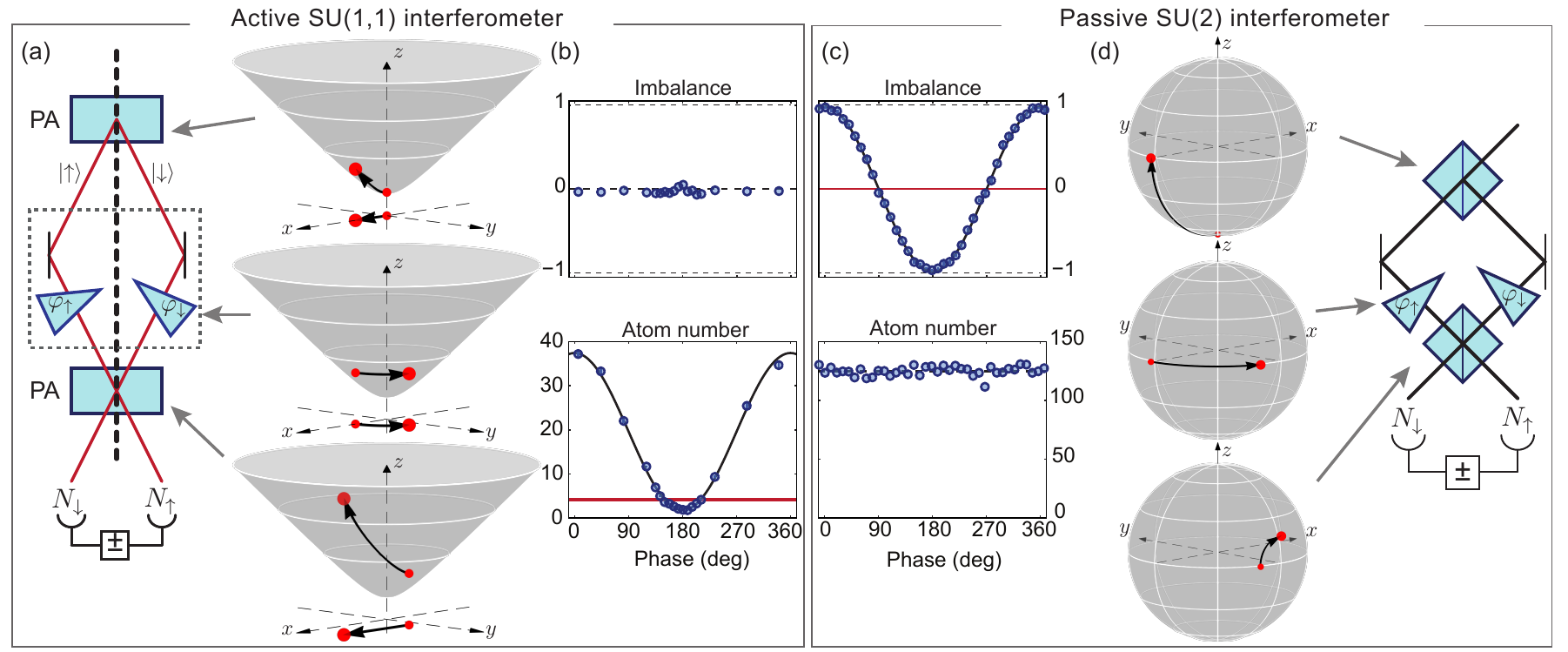}
\caption{\textbf{Active atom interferometry} \textbf{(a)} Schematic representation of an SU(1,1) interferometer, whose components' action on the two modes $\ket{\uparrow}$ and $\ket{\downarrow}$ can be visualised on hyperbolic surfaces (see text). Starting from vacuum the two modes (red, solid lines) get spontaneously populated by the first parametric amplifier (PA), which is driven by a strong pump field (dashed). On the hyperbolic surface, this process is represented by a boost along the $x$-axis (top). Phase accumulation by $\varphi_\downarrow+\varphi_\uparrow$ is represented by a rotation about the $z$-axis (middle), before a second PA performs a final boost, translating the phase into an either further amplified or deamplified population of the two modes (projection onto $z$-axis, i.e. $\braket{K_z}$) (lowest hyperbolic surface).
\textbf{(b)} Readout. All elements conserve the magnetization, i.e. the population imbalance remains vanishing for all phases. The phase-dependent fringe is recorded in the total number of particles leaving the interferometer, $N_\uparrow+N_\downarrow$. This is in contrast to a passive interferometer, shown in \textbf{(c)}. Here passive beam splitters redistribute the overall ever fixed number of atoms (lower panel) between two modes such that the interferometry fringe is recovered in atom number imbalance (upper panel). \textbf{(d)} Schematic representation of a Mach-Zehnder interferometer, prototypical for passive interferometry. The element's actions are illustrated on Bloch-spheres. Passive beam splitting constitutes rotations about the $y$-axis (top and lower panel), while phase accumulation by $\varphi_\uparrow-\varphi_\downarrow$ is represented as a rotation about the $z$-axis.}
\end{figure*}

We now discuss the interferometric sequence for both, active and passive beam splitting. For this we consider the two modes $\ket{\uparrow}$ and $\ket{\downarrow}$ to be populated by indistinguishable bosons. Passive interferometers can be described in a pictorial manner by resorting to the Bloch-sphere representation. In a similar fashion the sequence of an active interferometer can be described on a series of hyperbolic cones. Here, we will review both representations. 

\subsection{Active SU(1,1) Interferometer}
To describe the active interferometer and parametric amplification we introduce three operators \cite{Yurke1986}; 
\begin{equation}
\hat{K}_z = \frac{1}{2}(\hat{a}_\uparrow^\dagger\hat{a}_\uparrow + \hat{a}_\downarrow^\dagger\hat{a}_\downarrow + 1) = (\hat{N}_\uparrow+\hat{N}_\downarrow+1)/2
\end{equation}
encodes the number of atoms shared in both modes. 
\begin{equation}
\hat{K}_x = \frac{1}{2}(\hat{a}_\uparrow^\dagger\hat{a}_\downarrow^\dagger + \hat{a}_\uparrow\hat{a}_\downarrow) \; \text{and} \; \hat{K}_y = \frac{1}{2\ii}(\hat{a}_\uparrow^\dagger\hat{a}_\downarrow^\dagger - \hat{a}_\uparrow\hat{a}_\downarrow)
\end{equation}
describe the coherent and pairwise creation and destruction of particles. These three operators belong to the SU(1,1) group \cite{Schumaker1985,Yurke1986} as they satisfy its defining commutation relations, viz: 
\begin{equation}
[\hat{K}_x, \hat{K}_y] = -\ii \hat{K}_z, \; [\hat{K}_y, \hat{K}_z] = \ii \hat{K}_x, \; [\hat{K}_z, \hat{K}_x] = \ii \hat{K}_y. 
\end{equation}
The conserved quantity (Casimir invariant) of this group is $\hat{K}_\text{tot}^2 = \hat{K}_z^2-\hat{K}_x^2-\hat{K}_y^2$ which is related to the ever fixed atom number imbalance.

In geometric terms $\hat{K}_\text{tot}^2$ defines a hyperbolic surface (cone) in the space spanned by the $K_i$. Therefore, the states satisfying the SU(1,1) algebra can be visualised on the surface of a cone \cite{Yurke1986}. The vacuum state is represented at the bottom of this cone. Because within the undepleted pump approximation the number of atoms in modes $\ket{\uparrow}$ and $\ket{\downarrow}$ is unbounded, the cone is open to the top (see Figure 2a). The operators $\hat{K}_x$ and $\hat{K}_y$ are generators for boosts along the $x$- and $y$- direction \cite{Han1988,Baskal2004}, respectively, while $\hat{K}_z$ generates rotations about the $z$-axis \cite{Yurke1986}. 

The interferometric sequence consists of two SU(1,1) boosts in succession with a phase rotation in-between. Such a scheme is schematically drawn in Figure 2a (top to bottom) where the boxes represent parametric amplification (PA) of modes $\ket{\uparrow}$ and $\ket{\downarrow}$ (red) with the aid of a pump mode (dashed). This sketch draws analogy to quantum optics, where parametric down-conversion is similar to spin exchange in atom optics. Each component's action is represented on the hyperbolic surface: 

Starting from vacuum, a first pulse performs a boost along the $x$-axis (see trajectory projection underneath the cone). Hereby, the two modes become populated (projection onto $K_z$-axis). Subsequent phase evolution rotates the state about the $z$-axis, before a second boost of equal strength along the $x$-axis is applied. This readout boost translates the accumulated phase onto the $z$-axis. The interferometry fringe is thus obtained by measuring $\braket{\hat{K}_z}$. In terms of side mode population the interferometry fringe is given by $\braket{N_\uparrow+N_\downarrow}\propto1+\cos(\varphi_\uparrow+\varphi_\downarrow)$ (lower panel, Fig 2b) from which $\varphi_\uparrow+\varphi_\downarrow$ can be deduced. For a phase of $\pi$ the second boost reverses the effect of the first and the initial vacuum state is recovered at the output. In contrast, at the fringe's maximum both boosts act in the same direction and a large mean atom number is detected at the output as a consequence of nonlinear amplification. The projection of $\hat{K}_z-1/2$ after the first amplification, thus within the interferometer, is indicated by the red horizontal line. For all phase settings the atom number imbalance remains vanishing (upper panel, Fig 2b).

Instead of using two boosts along identical directions (as illustrated in Fig 2a) a similar interferometry fringe is obtained by changing the relative phase $\varphi_\text{ref}$ of the two boosts (without phase interrogation in-between). If the first boost acts along (say) $x$-direction such a phase change corresponds to the substitution of $\hat{K}_x\rightarrow\cos{\varphi_\text{ref}}\hat{K}_x+\sin\varphi_\text{ref}\hat{K}_y$ for the second boost. 

\subsection{Passive SU(2) Interferometer}

To describe the ubiquitous passive interferometer we resort to the SU(2) group description. As a prominent example, Ramsey's method of separate oscillatory fields falls into this class. This is the atomic analogue of the Mach-Zehnder interferometer known in optics (schematically drawn in Fig 2d) \cite{Lee2002}. 

We introduce pseudo-spin operators whose $z$-direction is given by the atomic population difference 
\begin{equation}
\hat{J}_z = (\hat{a}_\uparrow^\dagger \hat{a}_\uparrow - \hat{a}_\downarrow^\dagger \hat{a}_\downarrow)/2 = (\hat{N}_\uparrow - \hat{N}_\downarrow)/2.
\end{equation}
The corresponding coherences, i.e., $x$- and $y$-direction of the collective spin are given by 
\begin{equation}
\hat{J}_x = (\hat{a}_\uparrow^\dagger \hat{a}_\downarrow + \hat{a}_\downarrow^\dagger \hat{a}_\uparrow)/2 \; \text{and} \; \hat{J}_y = (\hat{a}_\uparrow^\dagger \hat{a}_\downarrow - \hat{a}_\downarrow^\dagger \hat{a}_\uparrow)/2\ii.
\end{equation}
These spin operators satisfy the defining commutator relation of the SU(2) algebra, i.e. $[\hat{J}_x, \hat{J}_y] = \ii \hat{J}_z$ and cyclic permutations thereof. 

The conserved total number of particles $N$ is encoded in the fixed spin-length $\hat{J}_\text{tot}^2 = \hat{J}_x^2+\hat{J}_y^2+\hat{J}_z^2$ which defines the surface of a sphere. As the angular momentum operators are generators of rotations, the action of these can be visualised on a generalized Bloch-sphere.

The interferometric sequence of such an SU(2) interferometer is built up by three rotations (shown in Fig 2d from top to bottom). Starting with all atoms in state $\ket{\downarrow}$ (south pole of the Bloch sphere), a $90^\circ$ rotation about the $y$-axis is performed to generate a phase sensitive linear superposition of $\ket{\uparrow}$ and $\ket{\downarrow}$. In this state the collective spin is aligned in the equatorial plane. Subsequent phase accumulation of $\varphi_\uparrow-\varphi_\downarrow$ is described by a rotation about the $z$-axis. Finally, a readout rotation of $90^\circ$ about the $y$-axis maps the accumulated phase onto the $J_z$ axis. By measuring the imbalance $2\braket{\hat{J}_z}/N=\braket{N_\uparrow-N_\downarrow}/N=\cos(\varphi_\uparrow-\varphi_\downarrow)$ the phase can be inferred (top, Fig 2c). The state inside the interferometer has a vanishing imbalance (horizontal red line). 

Similar to the SU(1,1) case, rotations about different axes can be implemented by phase changes $\varphi_\text{ref}$ of the linear coupling. When the first pulse rotates about the $y$-axis this phase change amounts to the substitution of $\hat{J}_y\rightarrow\cos\varphi_\text{ref}\hat{J}_y+\sin\varphi_\text{ref}\hat{J}_x$ for the final rotation.

\subsection{Coherent states: Fluctuations}
The SU(1,1) coherent states, i.e. those states which are generated by linear combinations of the $\hat{K}$ operators acting onto vacuum, are given by $\ket{r,\varphi}=\ee^{-\ii \varphi\hat{K}_z} \ee^{-\ii r \hat{K}_x} \ket{0}_\uparrow \ket{0}_\downarrow$. This can be understood as parametric amplification of vacuum fluctuations \cite{Leslie2009,Klempt2010,LewisSwan2013}. These states satisfy $\braket{\hat{K}_z}=\cosh(2r)/2$. In terms of Fock states they are represented by
\begin{equation}
\ket{r,\varphi} = \frac{1}{\tanh{(r)}} \sum_{n=0}^\infty e^{\ii n \varphi} \cosh^n{(r)} \ket{n}_\uparrow \ket{n}_\downarrow
\end{equation}
This coherent superposition of twin-Fock states $\ket{n}_\uparrow \ket{n}_\downarrow$ is a highly entangled state as witnessed by this Schmidt decomposition. In quantum optics this state, known as two-mode squeezed vacuum, is the archetype of an entangled state and is routinely generated by parametric down conversion \cite{Braunstein2005}. The individual modes do not have a mean field, i.e. $_\uparrow\bra{n}\hat{a}_\uparrow\ket{n}_\uparrow=0$ and similarly for mode $\ket{\downarrow}$ \cite{Haine2015}. Therefore, the mode's fluctuations become crucial: These characteristic fluctuations are thermal, meaning that compared to the mean mode populations, $\braket{N_\uparrow+N_\downarrow}\equiv\braket{N}$ the variance reads $\var{N}=\braket{N}(\braket{N}+2)$. 

The SU(2) coherent states are generated by the $\hat{J}$ operators acting onto the lowest lying state: $\ket{\theta,\varphi}=\ee^{-\ii \varphi \hat{J}_z} \ee^{-\ii \theta \hat{J}_y}\ket{0}_\uparrow\ket{N}_\downarrow$. With this operation each atom is independently put into the same superposition state: $\ket{\theta,\varphi}=(\cos\theta/2\ket{\downarrow}+\ee^{\ii\varphi}\sin\theta/2\ket{\uparrow})^{\otimes N}$. When measuring, each of the $N$ atoms is independently distributed among the two modes like in a Bernoulli trial. These states are therefore not particle entangled. Defining the success probability $p=(\cos\theta+1)/2$ a possible representation in Fock states reads
\begin{equation}
\ket{\theta, \varphi} = \sum_{n=0}^N e^{\ii n \varphi} \sqrt{ {{N}\choose{n}} p^n (1-p)^{N-n}} \ket{n}_\uparrow \ket{N-n}_\downarrow
\end{equation}
These SU(2) coherent states feature binomial fluctuations in population imbalance, i.e. $\var{(N_\uparrow-N_\downarrow)}=4p(1-p)\braket{N}$, as well as individual mode populations. For equal populations of both modes we have $\var{(N_\uparrow-N_\downarrow)}=\braket{N}$. 

This stark difference between binomial and thermal-like mode fluctuations is the subject of Figure 1.  

\subsection{Phase dependence}

The two classes of interferometers are sensitive to different phases. This is a consequence of the contrary conservation laws: A fluctuating total number of particles is necessary to define the sum phase $\varphi_+=\varphi_\uparrow+\varphi_\downarrow$. In turn, a fluctuating number imbalances is needed to define the difference phase $\varphi_-=\varphi_\uparrow-\varphi_\downarrow$. Consequently, the SU(1,1) interferometer is sensitive to the sum phase while the difference phase is read out in the SU(2) interferometer. 

All interferometers compare the phase accumulated during interrogation to the phase reference $\varphi_\text{ref}$ provided by the respective beam splitting mechanism. In a more complete picture of SU(1,1) interferometry, the relative phase shifts of parametric amplification (i.e. changes in boost direction) and the phase acquired during interrogation $\varphi_+$ are combined. The output fringe is then given by $\propto1+\cos(\varphi_\text{ref}-\varphi_+)$. Similarly, SU(2) interferometry yields a fringe $\propto\cos(\varphi_\text{ref}-\varphi_-)$.

\begin{figure*}
\includegraphics{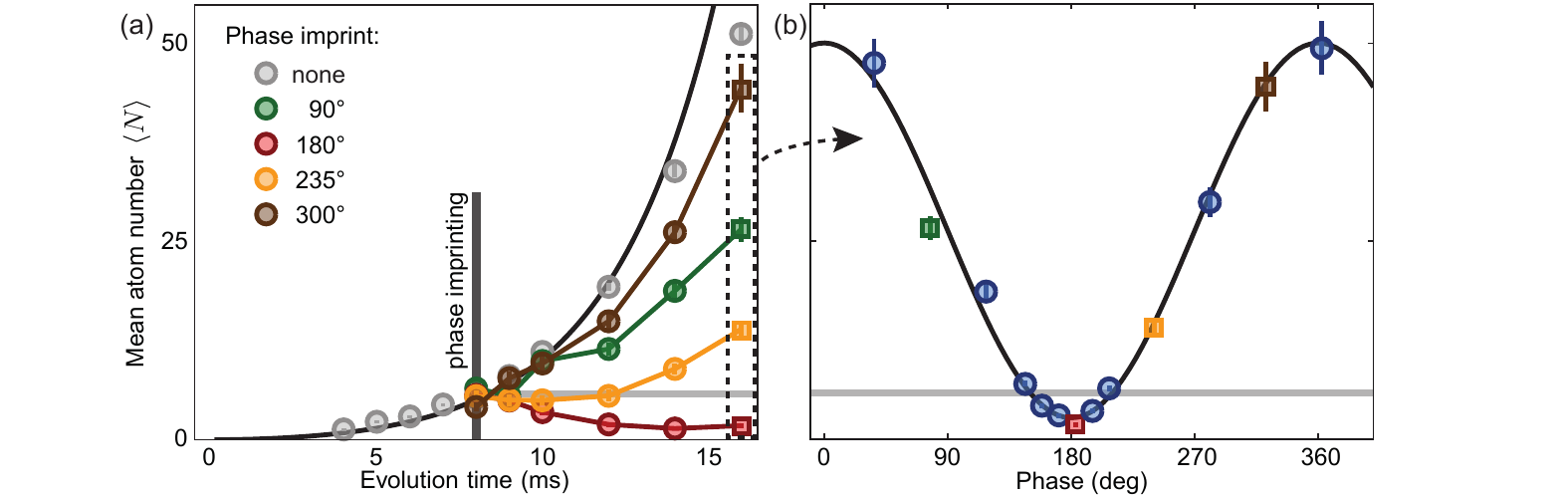}
\caption{\textbf{Time trace of mode population $\braket{N}$ during the interferometric sequence.} \textbf{(a)} We perform resonant spin-changing collisions for $t_1 = 8\,$ms thereby realising the first SU(1,1) boost. Subsequently, a phase shift $\varphi_\text{ref}$ is imprinted. For phase shifts close to $\varphi_\text{ref}\sim\pi$ the second period of spin-exchange is reversed (red trace) while for phases $\varphi_\text{ref} \sim 0$ the population $\braket{N}$ continues to grow (brown). \textbf{(b)} The characteristic fringe is obtained after having completed the second period of spin-exchange with $t_2=t_1$. The grey horizontal line indicates the average atom number $\braket{N}$ after the first boost. The readout boost starts from this level, clearly demonstrating the ensuing nonlinear amplification.}
\end{figure*}

\subsection{Hamiltonian in $\hat{K}$-representation}

Within the undepleted pump approximation our experimental system of a quasi spin-$1$ condensate is described by the Hamiltonian $\mathcal{H} = 2 \hbar \kappa \hat{K}_x + 2 \hbar \delta\hat{K}_z$. The detuning term $\delta=\kappa+qB^2+\delta_\text{MW}$ contains contributions of collisional shifts (via $\kappa$), Zeeman energy shifts ($qB^2$) and level shifts due to microwave dressing ($\delta_\text{MW}$). As magnetization conserving processes, all SU(1,1) transformations are unaffected by Zeeman energy shifts linear in magnetic field strength $B$. Only its quadratic contribution constitutes a detuning term $qB^2\hat{K}_z$ with $q\approx2\pi\times72\,$ Hz/$G^2$. To achieve resonance $\delta \approx 0$ we use dispersive microwave dressing mainly affecting the pump mode \cite{Gerbier2006}. 

By switching the phase of $\kappa\rightarrow\kappa\ee^{\ii\varphi_\text{ref}}$, the second boost acts along a different direction as described by the Hamiltonian $\mathcal{H} = 2 \hbar \kappa (\cos\varphi_\text{ref} \hat{K}_x + \sin\varphi_\text{ref} \hat{K}_y) + 2 \hbar \delta \hat{K}_z$. Experimentally, the phase $\varphi_\text{ref}$ of $\kappa$ is controlled by phase imprints onto the pump mode. Changing the pump phase by $\varphi_0$ yields twice the phase for $\varphi_\text{ref}=2\varphi_0$, which is a consequence of the pairwise scattering during parametric amplification.

\section{Experimental results}

\subsection{Experimental sequence}

We start with a Bose-Einstein condensate prepared in the $\ket{F,m_F}=\ket{1,0}$ state at a magnetic bias field of $B\approx0.9\,$G. To initialize the spin mixing we transfer all atoms to $\ket{2,0}$ by a microwave $\pi$-pulse. Thereby, the nonlinearity for parametric amplification is rapidly enhanced to $\kappa\rightarrow2\pi\times20\,$Hz. Subsequently, we use dispersive microwave dressing to fulfil the energy resonance $2\times\ket{2,0} \leftrightarrow \ket{2,1} + \ket{2,-1}$, i.e. $\delta \approx 0$. This realises a resonant SU(1,1) boost along the $x$-direction. Since the states $\ket{2,\pm2}$ are largely detuned due to the second order Zeeman shift ($2\pi\times180\,$Hz) and the associated coupling strength is weaker, the boost does not affect these states. To interrupt the boost at a particular point in time we stop applying microwave dressing and immediately transfer the pump atoms back to $\ket{1,0}$. 

In other experiments spin exchange is often performed within the $F=1$ manifold. Our reasons to use an effective three level system within the $F=2$ manifold instead are twofold; First, the coupling strength for spin mixing in $F=2$ is one order of magnitude larger than in $F=1$. This allows for more robust and faster spin dynamics. 

Secondly and more importantly, being able to quickly transfer the pump atoms into the $F=1$ state allows us to effectively stop the nonlinear dynamics. Since the detuning change by deactivating the microwave dressing is too small, residual off-resonant spin dynamics is still present. Therefore, the transfer of the pump mode into a state that does not participate in spin exchange is crucial for fast and precise control over the nonlinear coupling. With the pump mode transferred the population in $F=2$ is frozen. This is a result of the then diminished coupling strength for spin exchange and the correspondingly large detuning. 

When the pump mode is in $F=1$, spin dynamics is inhibited because the detuning $\delta\approx2\pi\times60\,$Hz is much larger than the relevant coupling strength for spin mixing within the $F=1$ manifold. Therefore the modes $\ket{1,\pm1}$ remain empty at all times and cannot act as a seed to classically speed-up spin exchange \cite{LewisSwan2013}. This scheme relies on the fact that the nonlinear coupling strength in $F=1$ is much smaller than for $F=2$. With this technique a purely linear phase evolution generated by $\hat{K}_z$ can be implemented. Without this shelving of the pump mode, nonlinear dynamics would still continue (albeit offresonant) and yield a combined evolution of $\hat{K}_x$ and $\hat{K}_z$ during the subsequent phase evolution.

For phase imprinting, we perform the microwave $\pi$-pulse used for shelving slightly detuned by $\delta_\text{MW}\approx2\pi\times100\,$Hz. Therefore, while having the pump mode in $F=1$ its phase evolves dynamically at a rate of $\delta_\text{MW}$. Holding the pump mode for time $t_\text{hold}$ before transferring it back to $F=2$ imprints the phase $\varphi_\text{ref}=2\delta_\text{MW}t_\text{hold}/\hbar$. The final boost is then implemented in $F=2$ similar to the first. 

The phase imprinting procedure can be described within the Bloch sphere representation that was introduced for the passive interferometry scheme. For this we identify the south and north pole with states $\ket{1,0}$ and $\ket{2,0}$, respectively. Then the first $\pi$-pulse (being only slightly detuned) rotates the state (almost completely) from the north to the south pole. During the subsequent hold time the state rotates about the $z$-axis at a rate given by the detuning of the microwave pulse, $\delta_\text{MW}$. The second microwave pulse rotates the state back to the north pole. Both $\pi$-pulses perform rotations about an identical axis on the Bloch-sphere. As a result, the initial and final state, both in $\ket{2,0}$ differ by a phase imprint. This phase imprint is composed of the dynamical phase $\delta_\text{MW}t_\text{hold}/\hbar$ and a geometric phase that is independent of the hold time. As we scan the phase via the hold time the geometric phase merely leads to a fixed phase offset. Experimentally we use holding times in $F=1$ of a few milliseconds, c.f. Figure 6. 

We first consider a single period of parametric amplification. While boosting from the vacuum state, the mean population grows nonlinearly described by $\braket{N_\uparrow+N_\downarrow}\equiv\braket{N}=2\braket{\hat{K}_z(t)}-1=2\sinh^2(\kappa t)$ (black line in Fig 3a). Experimentally, we find good agreement for evolution times ranging up to $\approx15\,$ms (grey points). For larger evolution times pump depletion successively reduces the coupling strength and thereby limits further growth. 

Additionally to this single boost, time traces of $\braket{N}$ during the entire interferometric sequence are shown in color (Figure 3a). Here, the first boost is stopped at $t_1=8\,$ms. By varying the holding time $t_\text{hold}$ of the pump atoms in $F=1$ (not shown in Figure 3a) a dynamical phase shift $\varphi_\text{ref}$ is imprinted as detailed above. After this holding time, the pump atoms are transferred back to $\ket{2,0}$ and microwave dressing is reapplied to realise the final SU(1,1) readout boost. 

For phases $\varphi_\text{ref}\approx\pi$ the two boosts act in opposite direction such that the population dynamics is reversed (red trace). For phase $\varphi_\text{ref}\approx0$ both boosts are in similar direction which is equal to the single resonant boost (grey). In the symmetric case of both boosts having the same duration $t_1=t_2$ (highlighted by the dashed box) the phase dependent fringe is recovered (Fig 3b) as indicated by the common plot markers. The solid line is a sinusoidal fit. Remarkably, compared to the average atom number $n$ inside the interferometer (indicated by the horizontal grey line) the size of the output fringe is nonlinearly enhanced, meaning that the fringe maximum is at $2n(n+2)$. 

\subsection{Fringe enhancement and noise suppression}

\begin{figure*}
\includegraphics{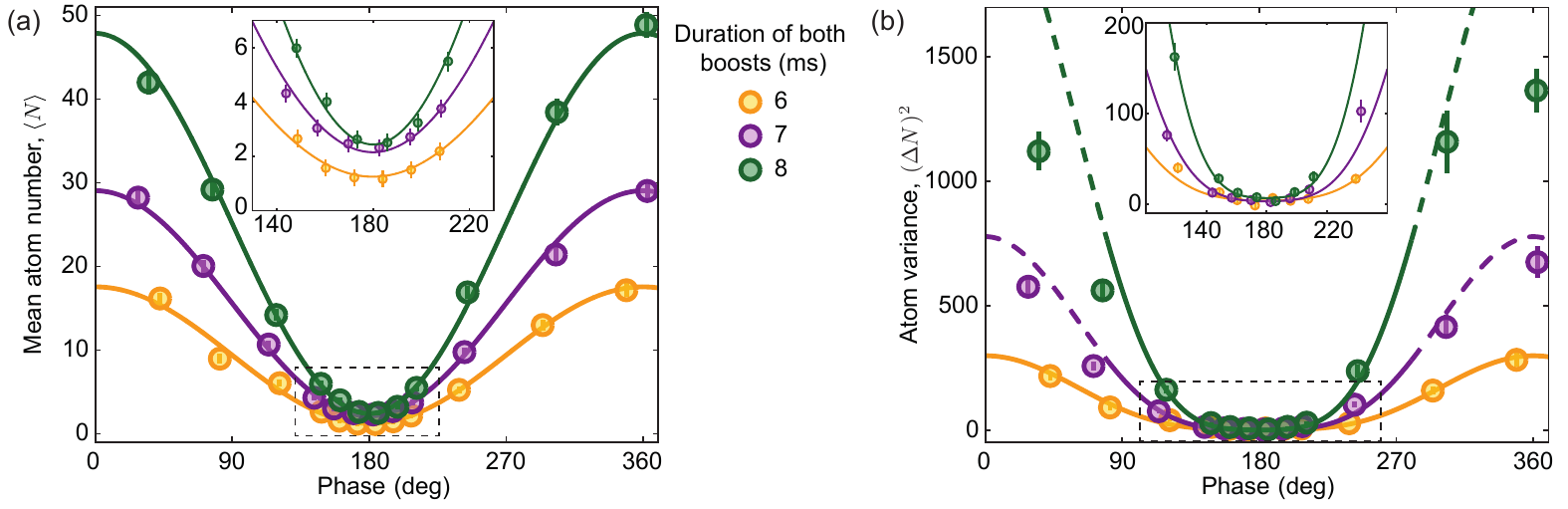}
\caption{\textbf{Fringe enhancement and noise suppression.} \textbf{(a)} Output fringes for the symmetric case of both SU(1,1) boosts having the same duration (indicated by colour). The atom number is nonlinearly amplified during the second SU(1,1) boost yielding strongly increased fringe heights. Solid lines are sinusoidal fits. Suppression at the minimum (inset) works best for short evolution times. \textbf{(b)} The variance shows a non-sinusoidal phase dependence with a flattened behaviour around the minimum at phase $\pi$ (inset). Solid lines are theory predictions within the undepleted pump approximation. For long evolution times pump depletion limits the variance growth at the maxima (dashed lines). }
\end{figure*}

The fringe enhancement is investigated by symmetrically altering the duration of both SU(1,1) boosts between $6-8\,$ms as shown in Figure 4. Increasing the evolution times we find dramatically magnified output fringes. Ideally, at the minimum perfect reversibility to vacuum is expected for all evolution times (within the undepleted pump approximation). Experimentally, however, we find a residual average population of $\approx0.5-1.5$ atoms per mode (see inset into dashed area) which is most likely caused by a non-perfect matching of both SU(1,1) boosts (see below). For the shortest evolution time the reversibility to vacuum is best. 

The measured fluctuations $\var{N}$ of the phase dependent output signal are shown in Fig 4b. The excess variance is distinctive for the SU(1,1) coherent states. Its theory prediction, $\var{N}=\braket{N}(\braket{N}+2)$ is indicated by the solid lines. These predictions take as input only the fitted mean fringe of panel a. For the largest evolution time the deviations at the maxima (dashed) result from pump depletion. As a function of phase $\varphi_\text{ref}$, the relationship between mean fringe and the corresponding variance leads to a non-sinusoidal dependence; the variance fringe consists of two Fourier components $\propto \cos\varphi_\text{ref}$ and $\propto \cos^2\varphi_\text{ref}$ yielding a flattened behaviour in vicinity of the minimum (inset into dashed area) \cite{Linnemann2016}. The contribution of detection noise is independently measured ($\approx33\,$ atoms$^2$) and is subtracted.  

\subsection{Phase sensitivity}

The interplay between fringe amplification and suppressed fluctuations at the minimum (dark fringe) allows for precise phase measurements. The phase sensitivity of the interferometer can be extracted by Gaussian error propagation on the mean atom number, $\var{\varphi} = \frac{\var{N}}{|dN/d\varphi|^2}$ \cite{Yurke1986, Marino2012}. To compare the phase sensitivity of this SU(1,1) interferometer with the typical reference of passive SU(2) interferometer, one has to define a common measure of the resource. Phase sensitivity is usually expressed in terms of the phase sensing mean atom number $\braket{N}$ \cite{Giovannetti2006, Pezze2015}; Therefore, only the atoms experiencing the phase shift $\varphi_+$ are counted as a an expensive resource. Classical probe states allow measuring at best at the so-called Standard Quantum Limit, i.e. $\var{\varphi}=1/\braket{N}$ \cite{Giovannetti2006}. The SU(1,1) interferometer allows surpassing this limit by exploiting the entangled SU(1,1) coherent states as demonstrated in \cite{Hudelist2014, Linnemann2016}. Best phase sensitivity is found in close vicinity to $\varphi=\pi$. Ideally, at the dark fringe a phase sensitivity at the ultimate Heisenberg limit is predicted, $\var{\varphi}=(\braket{N}(\braket{N}+2))^{-1}$. 

The amplification of the readout boost magnifies the signal, thereby generating larger slopes of the phase sensitive output signal $d\braket{N}/d\varphi$. However, posteriori amplification of classical states cannot enhance phase sensitivity since it does not differentiate between signal and noise and treats both likewise \cite{Clerk2010}. In contrast, the SU(1,1) interferometer features an improved signal-to-noise ratio because it harnesses both, the entanglement present at the probe stage generated by the first boost, as well as the second further (de-)amplifying boost \cite{Marino2012}. Best phase sensitivities are reached for matched durations of both boosts \cite{Linnemann2016}. 

\subsection{Non-balanced interferometry}

\begin{figure*}
\includegraphics{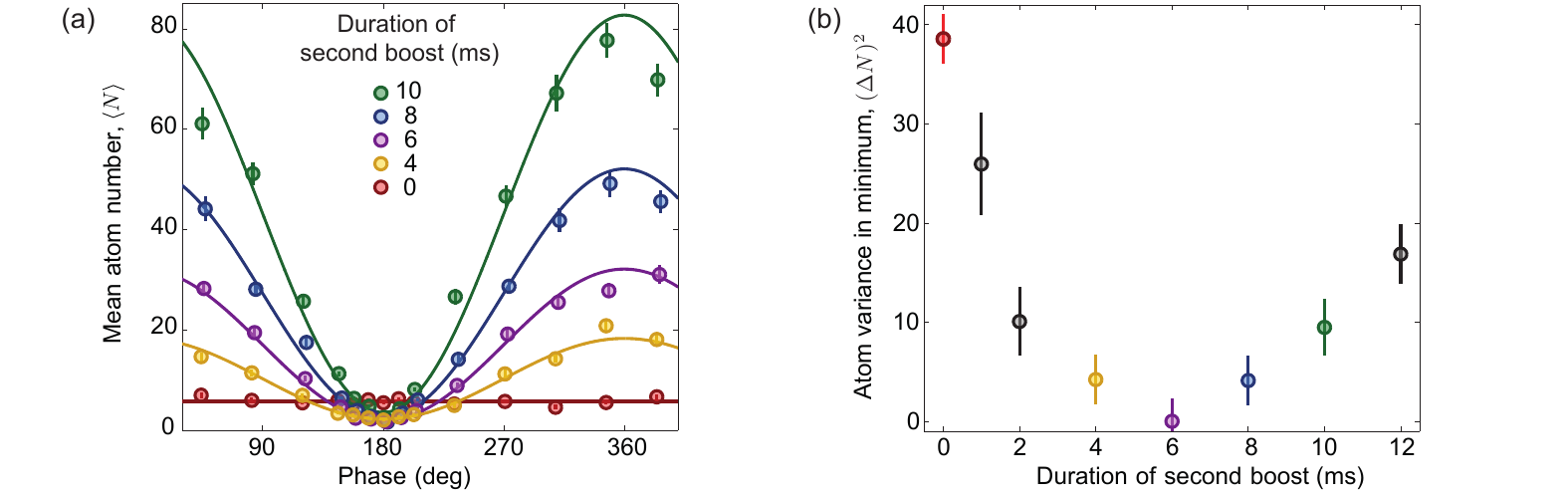}
\caption{\textbf{Suppression of fluctuations in the non-symmetric interferometer}. \textbf{(a)} Interferometry fringes for different durations of the second SU(1,1) boost while having the first fixed at $t_1=8\,$ms. The red data points show the outcome when the readout boost is omitted. Longer readout pulses $t_2>t_1$ amplify the output signal even further than the symmetric case. This amplification comes at the expense of incomplete suppression in the minimum. \textbf{(b)} Fluctuations at the fringe minimum for different durations of the readout boost. We find a pronounced minimum close to the symmetric case $t_1\approx t_2$ but shifted towards shorter durations for the second boost.}
\end{figure*}

The improved phase sensitivity relies crucially on the entanglement-enabled suppression of fluctuations \cite{Pooser2009, Kong2013} by the second SU(1,1) boost. To investigate this, we keep the first boost's duration fixed at $t_1=8\,$ms and vary the length of the readout boost. Perfect reversibility is only expected in the symmetric situation when both boosts have equal duration such that complete cancellation is reached at phase $\pi$. Figure 5a shows the fringes of $\braket{N}$ for unsymmetrical evolution times of $t_1=8\,$ms and the second boost ranging between $t_2=4-10\,$ms. By omitting the second boost entirely we probe the state inside the interferometer (shown in red) which shows no phase dependence as expected. Based on this state, the final boost either amplifies or absorbs the population in modes $\ket{\uparrow}$ and $\ket{\downarrow}$. Using longer readout pulses $t_2>t_1$ (green curve) amplifies the output fringe even further than the symmetric case (blue). However, this fringe enhancement comes at the expense of incomplete suppression in the minimum. This becomes most visible when considering the fluctuations at the minimum versus boost duration (Fig 5b): Optimal reduction of the fluctuation is reached when both boosts compensate each other, requiring $t_1 \approx t_2$. Experimentally, we find the best reduction of fluctuations for shorter times of the second period. Although it is unclear what causes this, it partially accounts for the residual population found in the minimum of the fringes which were obtained with symmetric durations as discussed in Figure 4a. 

\subsection{Sensitivity to collisional shifts}

The detuning term $\delta$ contains collisional shifts via $\kappa=gN_0$. Therefore, the nonlinear coupling strength not only parametrises the strength of the SU(1,1) boosts but also implies energy shifts during phase interrogation. By choosing experimental runs with vastly different pump atom numbers $N_0$ we can quantify this collisional shift and thereby measure the nonlinear coupling strength $\kappa$ as a function of atom number. Figure 6a shows the SU(1,1) interferometry fringes obtained when post-selecting atom numbers between $N_0=300$ (red) and $N_0=600$ (pink). The fringes dephase by $\approx2\pi$ after holding times exceeding $50\,$ms (right panel of Fig 6a). Fitting the frequency of the fringes shows a square root like atom number dependence (Figure 6b). This behaviour is expected for a trapped mesoscopic Bose-Einstein condensate for that the mode function overlap $\int \! \mathrm{d}^3x |\Psi(\bm{x})|^4$ lies in a cross over regime between that for single-particle wave functions and that within the Thomas-Fermi approximation. While in the former case interactions are neglected which leads to a constant mode overlap such that $\kappa\propto N$, the latter neglects kinetic energy resulting in a mode overlap $\propto N^{-3/5}$ leading to $\kappa\propto N^{2/5}$. 

\subsection{Detection requirements}

The fluctuations in readout direction ($\hat{K}_z$) are at all stages of the SU(1,1) interferometer super-Poissonian. Furthermore, the SU(1,1) coherent states are completely determined by their mean atom number corresponding to the single boost parameter $r$ of the SU(1,1) coherent states. Since the states feature broad histograms without any detailed finer structure, the requirements on detector resolution are significantly relaxed. To leverage the improved phase sensitivity, only mean values and no higher moments need to be recorded at the detector \cite{Marino2012,Cheung2016}. 

This stands in contrast to using highly entangled states with linear readout. When, for instance, the two-mode squeezed vacuum state is fed into a passive SU(2) interferometer, single-atom resolved measurements such as parity detection need to be performed \cite{Anisimov2010}. For the atom number parity, the single shot outcome is assigned $+1$ when an even number of atoms is detected at one of the output ports, and $-1$ for an odd atom number, respectively. Such parity detection schemes are characterised by a threshold behaviour: If the detection noise is too large to determine the parity in a single shot, averaging many realisations cannot restore the output signal at all. The uncertainty obtained with the SU(1,1) interferometer, on the other hand, can be reduced by averaging no matter how large the detection noise is. This immunity towards detection noise \cite{Davis2016, Froewis2015, Macri2016, Hosten2016, Leibfried2004} is enabled by the final boost which constitutes a nonlinear readout. 

\begin{figure*}
\includegraphics{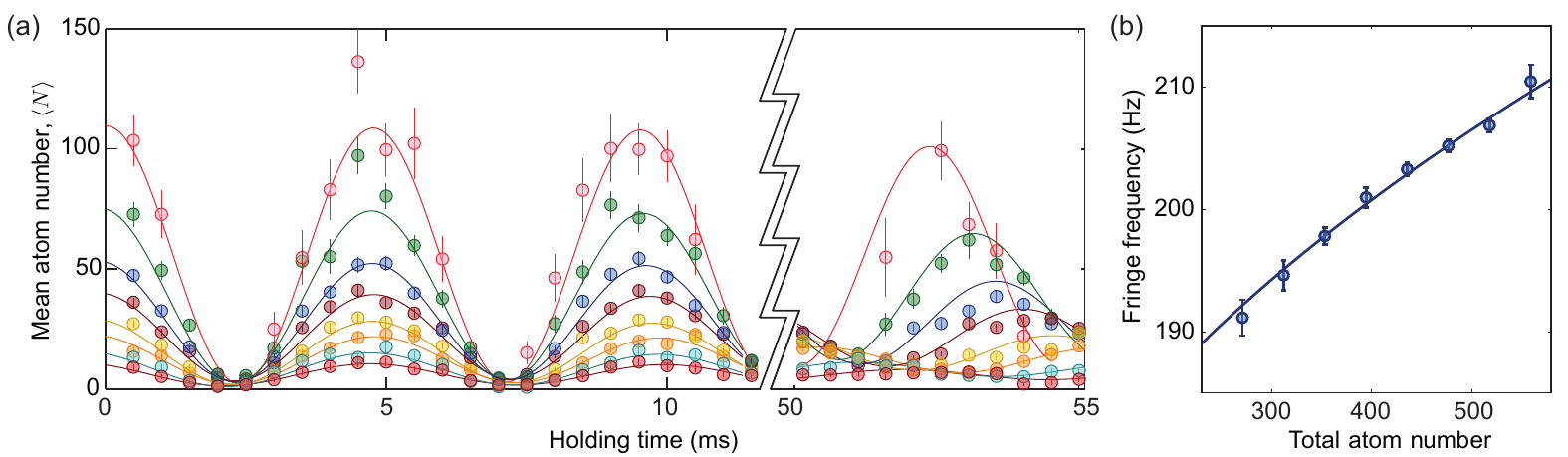}
\caption{\textbf{Collisional shifts during interrogation}. \textbf{(a)} The nonlinear coupling via spin-exchange implies energy shifts of the involved modes. These energy shifts constitute a detuning during phase interrogation. By post-selecting experiments with different total atom numbers we interferometrically measure the coupling strength $\kappa$ as a function of pump size. The pink interferometry fringes are obtained for choosing $N_0=600$ while the red data belongs to pump sizes of $N_0=300$. During the holding time which ranges up to $55\,$ms (right panel) the pump atoms are shelved in $F=1$ and almost no reduction of contrast occurs. \textbf{(b)} Observed frequency of the interferometry fringes as a function of total atom number. The solid line is a square root. }
\end{figure*}

\section{Conclusion}

We detailed the first demonstration of an active interferometer with atoms. From a more general point of view, the interferometric sequence constitutes a nonlinear time reversal sequence \cite{Linnemann2016} which is a particular useful form of nonlinear readout \cite{Froewis2015, Davis2016, Macri2016, Hosten2016}. Due to its high degree of experimental control we believe that this nonlinear readout can be used to characterise and detect entanglement present in condensates involving many spatial modes \cite{Boyer2008}. 

Within the undepleted pump approximation a large portion of the atoms provides the nonlinear mechanism without being used for phase estimation. Surpassing the limitation arising due to the small fraction of sensing atoms has been the subject of recent theoretical investigations: Reference \cite{Gabbrielli2015} predicts improved phase sensitivities also when severely depleting the pump mode, while Ref. \cite{Szigeti2016} suggests to use a combination of passive and active beam splitting. 

For some applications, however, the uneven partitioning of atoms constitutes an advantage: Interferometry between the large pump mode and the sparsely populated but highly sensitive probe fields might allow to investigate fundamental questions about phase diffusion in Bose-Einstein condensates \cite{Sinatra2008}. 

The coupling strength for spin-exchange within the $F=1$ manifold has an inverted sign as compared to $F=2$. Therefore, by using one period of spin-exchange within $F=1$ and another within $F=2$ (after swapping all three involved states) the Hamiltonian can be inverted without relying on the validity of the undepleted pump approximation. This extended time reversal might allow to study out-of-time-order correlations \cite{Li2016, Garttner2016} and the associated scrambling of quantum information \cite{Swingle2016}. 

\section{Acknowledgments}
The authors are indebted to Robert J. Lewis-Swan and Karen V. Kheruntsyan. This work was supported by the Heidelberg Center for Quantum Dynamics, the European Comission FET-Proactive grant AQuS (Project No. 640800), and the European Research Council (ERC) under the European Union's Horizon 2020 research and innovation programme (Grant agreement No 694561)


%

\end{document}